# Ultralong quantum optical storage using reversible inhomogeneous spin ensembles with an optical locking method

Byoung S. Ham

*The Graduate School of Information and Communications, and Center for Photon Information Processing, Inha University, 253 Younghyun-dong, Nam-gu, Incheon 402-751, S. Korea*

A novel method of multi-bit quantum optical data storage is presented, where the storage time can be lengthened far beyond the spin phase-decay time in a reversible spin inhomogeneous system excited by consecutive resonant Raman optical data pulses. The ultralong storage time is obtained by an optical population locking mechanism of modified rephasing process. This gives potentials to quantum repeaters utilizing quantum memories for long distance quantum communications, in which ultralong storage time plays a major role.

Quantum information transfer in a scalable quantum network composed of remote nodes has been a key research topic in the area of quantum communications utilizing quantum repeaters [1-4]. For the quantum repeaters quantum memories plays an important role. For long distance quantum transmission with quantum repeaters [3,4], ultralong quantum optical storage is an ultimate condition. Recent advancement of reversible quantum mapping processes between light and matter shows feasibility of quantum memory applications using a collective atomic ensemble as a quantum node [5-8]. During the last several years, various types of quantum optical data storage technique have been introduced including off-resonant Raman scattering [5], optically reversible photon echoes [6], electro-optic echoes [7], and slow light based reversible quantum mapping [8]. All of the experimental demonstrations of quantum memories show a proof of principle. To date, the storage time has been limited by a short time scale < ms.

Here, we present a novel method of quantum optical data storage technique to lengthen the storage time in several orders of magnitude up to several hours, to support quantum repeaters working for remote quantum nodes, and to guarantee nearly perfect retrieval efficiency. A primitive model of the present technique has already been demonstrated in 1997 [9], but the physics has not been carefully discussed nor clearly understood. This Letter seeks to enlighten the physics of the previously demonstrated reversible optical data storage technique and to present a long-lasting reversible quantum mapping protocol for quantum memory applications.

The physics of the present quantum optical data storage technique lies in a resonant Raman optical field-excited dark state [10], where the dark state is an imprinted form of an optical signal in a collective spin ensemble [9,11].

Contrary to previously demonstrated quantum data storage methods using optical inhomogeneous broadening such as photon echoes [6,7], the present method utilizes spin inhomogeneous broadening via two-photon resonance (dark state) by the resonant Raman optical pulse, where the quantum optical information of consecutive resonant Raman optical pulses is stored in a spectrally chosen atomic ensemble as superposed spin coherence gratings. To retrieve the stored optical information, a resonant Raman rephasing pulse(s) is applied to the atomic medium, and then the stored spin coherence is retrieved in a time reversed manner resulting in consecutive dark state coherence burst. For the ultralong optical data storage, an optical population locking method is introduced. The optical population locking technique is essential to avoid the spin coherence leakage resulting from optically excited atoms in the rephasing processes.

Photon echoes flourished in 1980s for multidimensional optical data storage [12]. In a two-level system the photon (spin) echo technique uses inhomogeneously broadened optical (spin) transition of atoms (spins) to store consecutive optical (spin) data into reversible optical (spin) spectral gratings, where the consecutive optical (spin) data can be spectrally superimposed in the same group of atoms (spins). In 2001, a Swedish group suggested a photon-echo-based quantum memory utilizing backward retrieval geometry for near perfect retrieval efficiency [6]. Later an Australian group experimentally proved the Swedish idea with an electro-optic quantum memory technique [7].

The present quantum optical data storage method is similar to the photon echo techniques using reversible inhomogeneous broadening. However, it is fundamentally different from in using spin inhomogeneous broadening as a storage medium of the optical information. To excite spin coherence (dark state), a resonant Raman optical pulse composed of a signal pulse $\Omega_P$ and a coupling pulse $\Omega_C$ is used (see the inset of Fig. 1(a)), where the coupling pulse assists storing the probe information into the spin coherence until the probe light is fully absorbed by the medium. In this context the medium must be optically dense enough for the probe to be absorbed completely.

For the numerical calculations, the spin inhomogeneous broadening is equally divided by a 2 kHz space, and time-dependent density matrix equations are numerically solved under the rotating wave approximation. The density matrix approach is very powerful in dealing with an ensemble system interacting with coherent laser fields owing to statistical information as well as quantum mechanical information [13]. The equation of motion of the density matrix operator ρ is determined from Schrödinger's equation [13]:

$$\frac{d\rho}{dt} = -\frac{i}{\hbar}[H,\rho] - \frac{1}{2}\{\Gamma,\rho\}, \quad (1)$$

where $\{\Gamma,\rho\}$ is $\Gamma\rho - \rho\Gamma$.

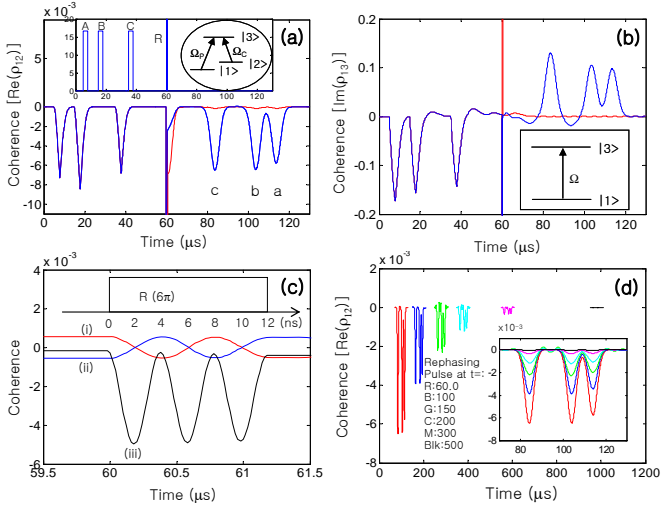

FIG. 1. Numerical simulations of a triple-bit quantum optical data storage based on reversible spin inhomogeneous broadening. Optical population (phase) decay rate from |3> to |1> or |2> is $\Gamma_{31}=\Gamma_{32}=0.5$ ($\gamma_{31}=\gamma_{32}=25$), spin population (phase) decay rate from |2> to |1> is $\Gamma_{21}=0$ ($\gamma_{21}=1$), spin inhomogeneous width between |1> and |2> is 200. For R, $\Omega_P=\Omega_C=50/\sqrt{2}$; for A, B, and C, $\Omega_P=\Omega_C=17$; initial population distribution is $\rho_{11}=\rho_{22}=0.5$. (a) a coded quantum optical data storage. The rephasing pulse area of R is $2\pi$ ($4\pi$) for the blue (red) curve. (b) photon echo (Im$\rho_{13}$) with an optical inhomogeneous width of 200 kHz: $\Gamma_{31}=0.5$; $\gamma_{31}=2.5$ ; $\Gamma_{32}=\gamma_{32}=0$; $\Gamma_{21}=\gamma_{21}=0$; $\rho_{11}=1$; $\rho_{22}=\rho_{33}=0$; $\Omega_P=25$; $\Omega_C=0$. (c) Coherence evolution of dark state (spin coherence $\rho_{12}$) by R (60.0<t<61.2 μs). (i) and (ii) Im$\rho_{12}$ for δ=−10 and δ=10, respectively, (iii) Re$\rho_{12}$ for δ=±10, (d) Spin decay-time dependent retrieval efficiency; R: red, B: blue, G: green, C: cyan, M: magenta, Blk: black. Unit: kHz.

Figure 1 shows numerical simulations of the present ultralong quantum optical data storage based on resonant Raman field-excited spin coherence (dark state) using a reversible spin inhomogeneous ensemble. The inset of Fig. 1(a) shows a schematic of a typical lambda-type three-level system interacting with resonant Raman optical fields composed of $\Omega_P$ and $\Omega_C$ at equal Rabi frequency. The upper figure of Fig. 1(a) represents the resonant Raman optical pulse sequence. The first three pulses stand for quantum optical data (A, B, and C; each pulse length is 3 μs), and the fourth stands for a rephrasing (R) Raman pulse. All of the medium's parameters are taken from experimental values of $Pr^{3+}$ doped $Y_2SiO_5$ (Pr:YSO) at liquid helium temperatures [9,14]. For simplicity, optical transitions are assumed to be homogeneous, where the width is effectively affected by the laser jitter [14]. The resonant Raman fields excite individual spins in the inhomogeneously broadened spin ensemble, where the spin inhomogeneous broadening is presumably assigned to the state |2> with a Gaussian distribution of 200 kHz (FWHM).

Figure 1(a) shows overall resonant Raman pulse excited spin coherence and retrieved signals by a $2\pi$ rephasing pulse R (blue curve). Less than 100% retrieval efficiency is due to the spin phase decay rate given by 1 kHz. On the other hand a $4\pi$ rephasing pulse R (red curve) produces almost zero retrieval efficiency (will be discussed below in Fig. 1(c)). The sequence of the retrieved signals (**c**, **b**, and **a**) is time reversed, which demonstrates a reversible spin inhomogeneous broadening mechanism as in photon echoes. The initial condition of population distribution on states |1> and |2> does not affect the retrieval efficiency if a balanced $2\pi$ rephasing pulse is applied (discussed elsewhere).

For comparison and validity of numerical calculations performed in Fig. 1(a), photon echoes are simulated in Fig. 1(b). With the same parameters used in Fig. 1(a) except the optical decay rate, the photon echo efficiency reveals zero as expected due to a complete phase recovery with a $2\pi$ rephasing pulse (see the red curve). The photon echo (retrieval) efficiency, however, shows a maximum with a π rephasing pulse (see the blue curve) due to complete phase reversal [12]. Thus, the validity of the quantum memory simulations in Fig. 1(a) is proven.

Figure 1(c) shows rephasing process for Fig. 1(a). The red curve (i) and the blue curve (ii) represent the imaginary part (Im$\rho_{12}$) of the dark state (spin coherence $\rho_{12}$) for a pair of symmetrically detuned spins at δ = 10 kHz and δ = −10 kHz, respectively. The black curve (iii) represents the corresponding real part: Re$\rho_{12}$. For the first $2\pi$ rephasing Raman pulse area (t = 4ns), the phase of the real part becomes recovered fully, while the phase of the imaginary part becomes completely reversed. The reversed phase of Im$\rho_{12}$ implies the time-reversal process of Fig. 1(a) similarly as in the π-pulse based photon echo in Fig. 1(b). For the $4\pi$ rephasing pulse area (t = 8ns), however, both real and imaginary parts recover their coherences fully, so there is no coherence retrieval (see also the red curve in Fig. 1(a)). This means that the rephasing process has modulo $4\pi$ in the rephasing pulse area. Thus, the $6\pi$ rephasing pulse (t = 12 ns) gives the same effect on the retrieval efficiency as the $2\pi$ rephasing pulse (not shown). In conclusion, for the maximum retrieval efficiency, the rephasing pulse area $\Phi_R$ should be:

$$\Phi_R = \int \Omega dt = 2\pi + 4n\pi, \quad (2)$$

where $\Omega$ is a generalized Rabi frequency of the resonant rephasing Raman pulses of $\Omega_P$ and $\Omega_C$, and n is an integer.

Figure 1(d) shows retrieval efficiencies depending on the $2\pi$ rephasing-pulse delayed at t=60 (red), 100 (blue), 150 (green), 200 (cyan), 300 (magenta), and 500 μs (black). The retrieval efficiency drops exponentially as a function of the delay time of the rephasing pulse R. The delay-time dependent retrieval efficiency fits into an exponential curve, $\exp(-t/T^S_2)$ as expected in the quantum mapping processes [5-9,11], where $T^S_2$ is spin phase decay time. As seen in the inset of Fig. 1(d), where the retrieved signals are intentionally overlapped, the retrieved signals have the same pulse shape regardless of retrieval efficiency. Thus, the data capacity of the present optical data storage is limited by the (spin) phase decay time $T^S_2$, where the storage capacity $N_Q$ is determined by the ratio of the data pulse length $\tau$ (3 μs) to $T^S_2$ ($T^S_2 = 1/(\pi\gamma_{21}) = 318$ μs).

We now suggest a novel method of ultralong quantum optical data storage limited by spin population decay time. Because the spin population decay time is much longer than the spin phase decay time in rare-earth doped solids, the optical storage time can be lengthened by several orders of magnitude higher. Like stimulated photon echoes [15], the rephasing pulse may be divided into two pulses to lock the spin phase decay process. However, in the three-level system of Fig. 1, the first part of the divided Raman rephasing pulse excites nearly 50% of atoms to the excited state $|3\rangle$. Then, the excited atoms decay down to the ground states considerably contributing to the dark state (spin phase) leakage. If the atoms on the excited state $|3\rangle$ can be safely removed to an isolated state and be delivered back later, then the dark state leakage can be frozen. We call this process optical population decay lock. For the optical population decay lock, an auxiliary spin state $|4\rangle$ is introduced with an auxiliary optical pulse $\Omega_A$ (see Fig. 2(a)). We assume a little bit ideal model whose spin transition rate between the existing ground states ($|1\rangle$ and $|2\rangle$) and the auxiliary state $|4\rangle$ is zero. The optical transition from states $|3\rangle$ to $|4\rangle$, however, is allowed similarly to the others.

To prove the present spin population decay time-limited quantum optical data storage, the rephasing pulse sequence is intuitively designed as shown in Fig. 2(b), where the first $\pi_R$ and last $3\pi_R$ pulses are for the resonant Raman pulses, whose pulse area is $\pi$ and $3\pi$, respectively. The auxiliary $\pi_A-\pi_A$ optical pulse sequence by $\Omega_A$ is provided to lock the optical population on state $|3\rangle$ via complete population transfer to states $|4\rangle$ and getting back to state $|3\rangle$. In this population transfer process, however, the overall dark state phase changes by 180 degrees as shown in Fig. 2c: See the $\pi$ phase change on each red and blue curve by the auxiliary $\pi_A-\pi_A$ optical pulses during $t_2$ (60.2 μs) $\leq t \leq t_5$ (1070.2 μs). This means that each imaginary component of the dark state (spin coherence, Im$\rho_{12}$) experiences an additional $\pi$ phase shift resulting in detuning swap between the symmetric spins: Im$\rho_{12}(\delta) \leftrightarrow$ Im$\rho_{12}(-\delta)$: The effect is similar to the DC electric field rephasing process in Ref. 7. Thus, total phase gain of the excited spin coherence (Im$\rho_{12}$) for $t_1 \leq t \leq t_5$ is $(3/2)\pi$ including $(1/2)\pi$ by the first Raman rephasing pulse $\pi_R$.

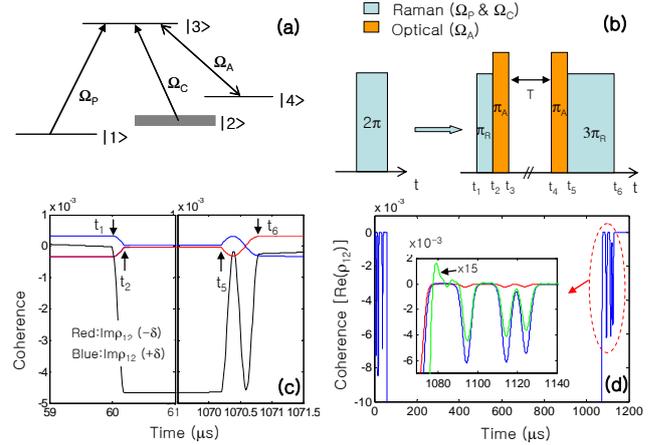

FIG. 2. Spin population decay-time limited quantum optical data storage. All simulation parameters are same as those in Fig. 1a. $\Omega_A$=50 kHz, population (phase) decay rate from $|3\rangle$ to $|4\rangle$, $|4\rangle$ to $|1\rangle$, and $|4\rangle$ to $|2\rangle$ are $\Gamma_{34}$= 0.5 ($\gamma_{43}$=25), $\Gamma_{41}$=0 ($\gamma_{41}$=0), $\Gamma_{42}$=0 ($\gamma_{42}$=0) kHz, respectively. (a) An energy level diagram. (b) Modified rephasing pulse sequence. (c) Coherence evolution of dark state (spin coherence $\rho_{12}$) by $\pi_R-\pi_A-\pi_A-3\pi_R$ rephasing pulse sequence at $t_1$=60.0, $t_2$=60.2, $t_3$=60.4, $t_4$=1070.0, $t_5$=1070.2, and $t_6$=1070.8 μs. (d) $\Omega_P$-invariant quantum optical memory whose storage time is much longer than spin phase decay time. Inset, Blue curve: $\Omega_P=\Omega_C$=16.7 kHz; green curve: $\Omega_P$=0.05 kHz and $\Omega_C$=25 kHz; red curve: with $\pi_R-\pi_A-\pi_A-\pi_R$ rephasing pulse.

The last Raman pulse $3\pi_R$ needs to induce $(3/2)\pi$ phase shift to make a complete phase reversal on the spin coherence as discussed in Fig. 1(c). Thus the pulse of $3\pi_R$ must be $3\pi$ as seen in Fig. 2(c) for $t_5 \leq t < t_6$. The temporal gap or storage time T (T=1010 μs) between two auxiliary optical $\pi_A-\pi_A$ pulses is long enough to be dephased completely if the system is for Fig. 1. In Fig. 2(d), however, the retrieval efficiency with $\pi_R-\pi_A-\pi_A-3\pi_R$ rephasing pulse sequence modification is same as that with a single $2\pi$ rephasing pulse in Fig. 1(a): See the blue curve in red-dashed circle and the inset for extension. Thus, the pulse area $\Phi_{2R}$ of the last Raman pulse should be:

$$\Phi_{2R} = \int \Omega dt = 3\pi + 4n\pi, \quad (3)$$

where $\Omega$ is generalized Rabi frequency of the resonant rephasing Raman pulse, and n is an integer. For the case of

$\Phi_{2R}=7\pi$, the same retrieval efficiency is obtained as in the case of $\Phi_{2R}=3\pi$ (not shown). For comparison, it should be noted that $\pi_R-\pi_A-\pi_A-\pi_R$ rephasing pulse sequence makes the retrieval efficiency severely drop almost down to zero (see the red curve in the inset of Fig. 2(d)). This is due to no spin-phase change during the rephasing process (see Fig. 2(c) for $t_1$ (60.0 μs) $\leq t < t_6'$ (1070.4 μs). The maximum storage time is thus determined not by spin phase decay time $T^S_2$ but by spin population decay time $T^S_1$. The only assumption made in the simulations is a zero spin-phase decay rate for $t_3 \leq t < t_4$ due to the result of the Block vector uv plane rotation by 90 degrees (see Ref. 13 and discussed elsewhere). In numerical simulations without using the auxiliary $\pi_A-\pi_A$ pulses, the retrieval efficiency is very low (not shown) as already shown experimentally in Ref. 9.

For potential applications of the present ultralong quantum optical data storage, extremely weak quantum optical signals such as single photons or a few photon entangled states may be considered for $\Omega_P$. Decreasing the Rabi frequency of $\Omega_P$ by a factor of 1/100 brings the retrieved value itself down to ~1/20 (see green curve in the inset of Fig. 2(d)), but the retrieval efficiency is nearly $\Omega_P$ value independent. In other words, the present method is robust even for extremely weak input signals of $\Omega_P$ compared to $\Omega_C$.

So far we have discussed a novel method of ultralong quantum optical data storage using reversible spin inhomogeneous broadening with an optical population decay locking method. To complete the quantum optical memory, however, the rephased spin coherence must be converted into optical fields. This spin-optical conversion process has been experimentally demonstrated using nondegenerate four-wave mixing processes [9,11] based on electromagnetically induced transparency (EIT) [16]. Because EIT makes an optically dense medium transparent to even resonant optical fields, the retrieved optical signal can be obtained without absorption. For the realization of quantum memory based on the present technique, population preparation ($\rho_{11} = 1$; $\rho_{22} = 0$) may be required to avoid any unwanted signal during the spin-optical coherence process. For an experimental demonstration, we will apply the present method to a $Pr^{3+}$ doped $Y_2SiO_5$, whose energy level configuration satisfies Fig. 2(a).

In conclusion, we have proposed a novel quantum optical data storage method using a reversible spin inhomogeneous broadening with an optical decay locking method. The spin population decay time-limited quantum optical data storage protocol gives potential to quantum communications utilizing quantum repeaters with an advantages of ultralong storage time up to several hours in a rare-earth doped solid, nearly perfect retrieval efficiency, and ultrafast data bit rate.

This work was supported by the Creative Research Initiative Program (Center for Photon Information Processing) by MEST via KOSEF, S. Korea.